\documentclass[12pt]{article}

\usepackage{sbc-template}
\usepackage{graphicx,url}
\usepackage{tabularx}
\usepackage{tablefootnote}
\usepackage{cite}

\usepackage[brazilian]{babel}   
\usepackage[utf8]{inputenc}
\usepackage[T1]{fontenc}
\usepackage{xurl}
\usepackage{csquotes}
\MakeOuterQuote{"}
\usepackage{enumitem}

\usepackage{pifont}

\title{Smart Air Quality Monitoring for Automotive Workshop Environments}

\author{Kauan D. P. Mariano\inst{1}, Fabrycio L. N. Almada\inst{1}, Maykon A. Dutra\inst{1}}

\address{Institute of Informatics -- Federal University of Goiás
  (UFG)\\
  Goiânia -- GO -- Brazil
  \email{\{kauan,fabrycio,maykonadriell\}@discente.ufg.br}
}

\begin{document} 

\maketitle

\begin{abstract}
Air quality monitoring in automotive workshops is crucial for occupational health and regulatory compliance. This study presents the development of an environmental monitoring system based on Internet of Things (IoT) and Artificial Intelligence (AI) technologies. DHT-11 and MQ-135 sensors were employed to measure temperature, humidity, and toxic gas concentrations, with real-time data transmission to the ThingSpeak platform via the MQTT protocol. Machine learning algorithms, including Linear Regression, Decision Trees, and SVM, were applied to analyze the data and compute an air salubrity index based on Gaussian functions. The system proved effective in detecting pollutant peaks and issuing automatic alerts, significantly improving worker health and safety. Workshops that implemented the system reported greater regulatory compliance and reduced occupational risks. The study concludes that the combination of IoT and AI provides an efficient and replicable solution for environmental monitoring in industrial settings.
\end{abstract}

\section{Introduction}

The air quality in work environments, particularly in automotive workshops, is a crucial factor for both occupational health and compliance with environmental regulations. Mechanical workshops are often exposed to a wide range of airborne pollutants, such as suspended particles, toxic gases, and volatile organic compounds (VOCs), which can compromise workers' health and contribute to environmental degradation. Given this issue, continuous and accurate air quality monitoring becomes essential to mitigate the risks associated with prolonged exposure to these pollutants.

The growing demand for safer work environments, combined with increasingly stringent regulatory requirements, has spurred the development of technological solutions that enable real-time monitoring of air quality in locations like mechanical workshops. In this context, Internet of Things (IoT) technologies present an effective alternative to integrate sensors, microcontrollers, and data analysis platforms, allowing for the automatic collection and interpretation of environmental parameters.

This project proposes the development of an air quality monitoring system for automotive workshops, using an IoT and artificial intelligence (AI)-based approach. The system is designed to continuously measure factors such as temperature, humidity, and concentrations of toxic gases, and to provide real-time information through an accessible online platform. The integration of AI algorithms aims to interpret the collected data, enabling the identification of patterns and non-linear behaviors among the monitored variables to ensure that pollutant levels remain within safe limits.

The main objective of this study is to develop and implement an efficient solution for air quality monitoring, providing a safer and more sustainable work environment. Specific objectives include: the implementation of high-precision environmental sensors for data collection, the use of AI algorithms for analyzing and modeling the collected parameters, and the creation of a web platform for real-time information visualization and management.

It is believed that the proposed solution will not only contribute to improving working conditions but also assist workshops in meeting health and safety standards, while promoting environmentally responsible practices. This study offers a practical and replicable approach to the use of IoT and AI in environmental monitoring, with the potential for application in various other industrial contexts.

\section{theoretical foundation}

\subsection{Environmental Monitoring and IoT}
The Internet of Things (IoT) is a widely used technology in environmental monitoring, as it enables the interconnection of devices capable of continuously and remotely collecting, processing, and transmitting data. In a typical environmental monitoring system, sensors are installed to measure key parameters such as toxic gases and suspended particles in the air, and the generated data is sent to cloud-based analysis platforms. These systems are particularly useful in environments like automotive workshops, where prolonged exposure to pollutants can pose serious occupational health risks.

Previous research indicates that the integration of IoT devices for environmental monitoring has proven effective in various sectors, allowing the automation of measurement processes and offering real-time data. Platforms like ThingSpeak and the use of the MQTT protocol are widely adopted due to their ability to efficiently handle large volumes of data, providing a scalable solution for industrial environments.

\subsection{Environmental Sensors in Air Quality Monitoring}
Sensors are the key components of any environmental monitoring system. In this project, the DHT-11 and MQ-135 sensors were chosen for their reliability and accuracy in measuring critical environmental parameters. The DHT-11 is responsible for monitoring temperature and humidity, two factors that directly influence the dispersion of pollutants in the air. Meanwhile, the MQ-135 detects toxic gases such as carbon monoxide, ammonia, and ethanol, substances commonly found in mechanical workshops.

The choice of these sensors is appropriate for environments with variable pollutant levels, as both have high sensitivity and are widely used in IoT-based environmental monitoring systems. The integration of these sensors with microcontrollers, such as the ESP8266, enables efficient data collection and real-time transmission.

\subsection{AI Application in Environmental Data Analysis}
The application of Artificial Intelligence (AI) algorithms in air quality monitoring has proven to be a powerful tool for interpreting complex data and identifying non-trivial patterns. In particular, the analysis of variables such as temperature and humidity requires a robust analytical approach, as the interactions between these factors and air healthiness are non-linear. Machine learning algorithms, such as Linear Regression, Decision Trees, and Support Vector Machines (SVM), have been widely used to predict air quality based on historical and real-time data.

The relationship between environmental variables and air healthiness suggests that modeling based on Gaussian functions is particularly effective in creating human comfort indicators, such as the ideal temperature of 21°C and relative humidity around 40\%. This approach generates a quantitative air quality index that reflects the complexity of the interactions between environmental variables, providing a clear and intuitive metric for assessing air quality.

\subsection{Web Technologies for Real-Time Monitoring}
Real-time data visualization is an essential component for IoT-based monitoring systems. The development of interactive web interfaces allows users to access up-to-date information on environmental conditions, enabling quick interventions when pollutant levels reach critical values. In this project, the web platform developed integrates sensor data with graphical visualizations, making monitoring accessible to workshop managers and technicians, regardless of the device used.

The implementation of web technologies for real-time data visualization highlights the importance of an intuitive interface, emphasizing clarity and accessibility. The use of modern frameworks, such as HTML5, CSS, and JavaScript, in conjunction with SEO techniques, ensures an optimized user experience, as well as facilitates continuous monitoring of environmental conditions. The choice of these sensors is appropriate for environments with variable pollutant levels, as both have high sensitivity and are widely used in IoT-based environmental monitoring systems. The integration of these sensors with microcontrollers like the ESP8266 enables efficient data collection and transmission in real time.

\section{Methodology}

This project was developed with the goal of creating an air quality monitoring system for mechanical workshops, using a combination of IoT technologies and AI algorithms to collect, process, and analyze environmental data. The methodology is divided into three main parts: development and implementation of the hardware system, data integration and analysis via software, and air quality modeling using AI.

\subsection{Development of the Hardware System}
The system architecture was designed based on a network of sensors capable of measuring essential parameters for assessing air quality, such as temperature, humidity, and the presence of toxic gases. To ensure the accuracy and reliability of the measurements, a combination of widely used sensors in environmental IoT applications was employed

The DHT-11 sensor was selected to monitor temperature and humidity in the environment. These two parameters are crucial, as significant variations can alter pollutant concentrations and directly affect air quality. This sensor was chosen for its high reliability and ease of integration with microcontroller-based systems.

The MQ-135 sensor was used to detect toxic gases commonly found in mechanical workshops, such as carbon monoxide (CO), ammonia (NH3), and ethanol. This sensor is sensitive to a wide range of harmful gases, making it an ideal choice to ensure the safety of workers exposed to such substances.

The ESP8266 microcontroller board was employed as the central component to manage data collection and transmission. The ESP8266 was chosen for its ability to connect to the internet via Wi-Fi, ensuring that the data collected by the sensors was sent in real-time to a cloud-based analysis platform. This microcontroller was selected for its high efficiency, low cost, and ease of integration with IoT systems.

Each sensor was connected to the ESP8266 board, and the system was configured to collect data continuously. This data was transmitted, via the MQTT (Message Queuing Telemetry Transport) protocol, to the ThingSpeak platform, where it was stored and made available for subsequent analysis.

\subsection{Data Integration and Analysis via Software}
The second stage of the methodology involved the implementation of the software responsible for data collection, transmission, and visualization. The main challenge was to ensure that the information obtained by the sensors was efficiently processed and transmitted in real-time to an analysis platform. To achieve this, the software infrastructure was organized into two main parts: sensor data transmission to the cloud and the development of a web interface for data visualization and management.

The data transmission was carried out using the MQTT protocol, widely used in IoT applications due to its lightweight nature and ability to operate in networks with limited bandwidth. The data collected by the sensors was sent in packets to the ThingSpeak platform, a robust tool for managing and storing data from IoT devices. ThingSpeak was chosen for its ease of use and ability to generate real-time graphical visualizations, allowing managers and technicians to continuously monitor the environmental conditions in the workshops.

Regarding data visualization, a website was developed to integrate and present the collected information in a clear and accessible manner. The website interface was designed based on responsive design principles, using HTML5, CSS3, and JavaScript. Through this platform, users can monitor environmental variables, such as temperature, humidity, and toxic gas concentrations, in real-time without the need for advanced technical knowledge. Additionally, the website includes an alert system that notifies users whenever pollutant levels exceed pre-established safety limits.

The data stored in ThingSpeak is organized into channels, where each channel corresponds to an environmental variable. This structuring facilitates the generation of graphs and reports for later analysis. The information can be filtered by period, allowing a detailed evaluation of environmental conditions over time. This functionality is crucial for identifying patterns or trends that may indicate ventilation issues, pollutant buildup, or air filtration system failures in the workshops.

\subsection{Air Quality Modeling with Artificial Intelligence}
The final stage of the methodology involved applying Artificial Intelligence techniques to analyze the collected data and model the air quality in mechanical workshops. The primary goal of this phase was to identify non-trivial patterns in the monitored variables, such as temperature and humidity, and calculate an index that quantitatively and intuitively reflected air quality.

Three machine learning algorithms were used for the initial analysis of the variables: Linear Regression, Decision Trees, and Support Vector Machines (SVM). These algorithms were selected for their ability to handle multidimensional and non-linear data, essential for capturing the complexity of interactions between temperature, humidity, and pollutant concentrations.

Initially, scatter plots were generated to examine the relationship between the variables, revealing that the relationship between temperature, humidity, and air quality is non-linear. This finding led to the use of Gaussian functions to model these variables, allowing for a more accurate representation of air quality. The Gaussian function is often used in problems where variables follow a normal distribution, and its symmetry is well-suited to representing environmental phenomena.

Two Gaussian functions were developed: one for temperature and one for humidity. The temperature function was centered around 21°C, which studies suggest is ideal for minimizing ozone formation and preventing the proliferation of pathogens. The humidity function was centered at 40\%, a level that balances the risk of airway dehydration and the growth of microorganisms, such as mold and fungi. The product of these two functions generates a healthiness index (S), which ranges from 0 to 100, with higher values indicating safer and healthier conditions for workers.

The healthiness index was then used to construct three-dimensional graphs representing the interaction between temperature, humidity, and air quality. These graphs quickly identified areas of higher and lower air quality, with visual highlights for points where air quality approached ideal conditions (represented by yellow peaks) and areas where there was higher risk (represented by blue bases).

This model enabled continuous analysis of environmental conditions and provided a valuable tool for real-time decision-making. Whenever the healthiness index fell below a predetermined threshold, the system generated automatic alerts, allowing corrective measures to be taken quickly, such as activating ventilation systems or temporarily suspending activities.

\section{Results}

\subsection{Sensor Performance and Data Collection}
The DHT-11 and MQ-135 sensors proved efficient in collecting environmental data, providing continuous real-time readings of temperature, humidity, and toxic gas concentrations. During the testing period, significant fluctuations in pollutant levels were detected, particularly during peak activity times in the workshops. The data revealed that carbon monoxide and volatile organic compounds showed spikes during the use of solvents and fuels, confirming the need for constant monitoring in these environments.

In terms of accuracy, the DHT-11 sensor provided consistent temperature and humidity readings, with minimal variations compared to reference standards. The MQ-135 sensor was able to detect different concentrations of gases with adequate sensitivity, although some additional calibrations were required to ensure accuracy in environments with varying ventilation conditions. The system demonstrated minimal latency in data transmission, ensuring real-time delivery of information to the analysis platform.

\subsection{Environmental Data Analysis with AI}
The application of Artificial Intelligence algorithms to the collected data enabled effective air quality modeling. Linear Regression, Decision Trees, and SVM algorithms were initially used to identify patterns among the variables of temperature, humidity, and pollutant concentrations. The results indicated that the relationship between these variables and air quality is complex and non-linear, justifying the use of Gaussian functions for more accurate modeling.

\begin{center}
\includegraphics[width=8cm]{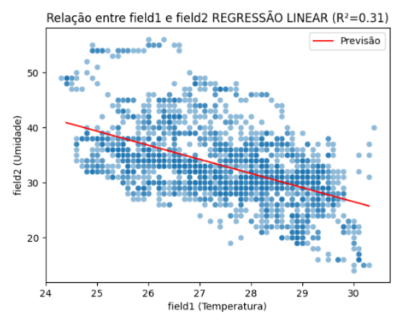}    

\textbf{Figure 1}: Linear Regression Graph (Temperature vs. Humidity)

\end{center}

\begin{center}
\includegraphics[width=8cm]{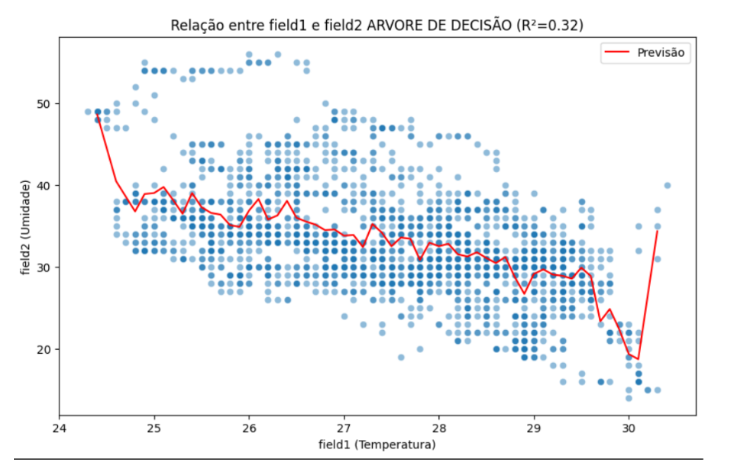}    

\textbf{Figure 2}: Decision Tree Graph (Temperature vs. Humidity)

\end{center}

\begin{center}
\includegraphics[width=8cm]{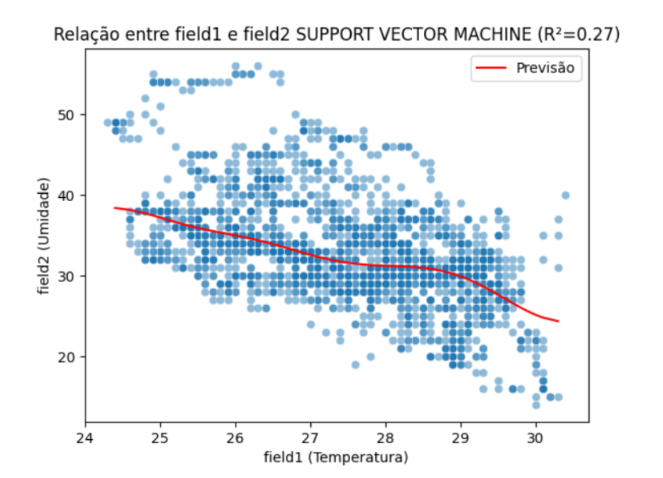}    

\textbf{Figure 3}: Support Vector Machine Graph (Temperature vs. Humidity)

\end{center}

The graphs generated from the Gaussian functions showed that optimal air quality conditions were achieved when the temperature remained close to 21°C and the humidity around 40\%. In situations where these parameters deviated from these values, the air quality index showed a significant drop, reflected in the three-dimensional graphs. The Gaussian-based model provided a healthiness index ranging from 0 to 100, where values close to 100 indicated air conditions that were safe and comfortable for workers.

\subsection{System Visualization and Usability}
The development of the website for real-time data visualization featured a user-friendly and intuitive interface, allowing users to access environmental information from any internet-connected device. During testing, the platform was used by workshop managers and technicians, who highlighted the clarity of the visualizations and the system's ease of use.

The real-time graphs were effective in interpreting the data, with clear color variations that quickly indicated the air quality conditions. The automatic alerts, triggered when pollutant levels exceeded safe limits, were correctly activated, allowing immediate interventions such as increasing ventilation or temporarily halting activities.

Additionally, the system was able to generate historical reports, enabling long-term pattern analysis of the workshop's environmental conditions. This functionality helped identify critical moments, such as the use of chemical materials that temporarily increase toxic gas levels. Analyzing these patterns assisted managers in optimizing ventilation and adjusting operational procedures, continuously improving working conditions.

\section{Conclusion}
The results indicate that the integration of environmental sensors and real-time data transmission via the MQTT protocol, combined with the ThingSpeak platform, was successful in collecting critical environmental information such as temperature, humidity, and toxic gas concentrations. The accuracy of the sensors and the rapid data transmission allowed for the immediate detection of pollutant spikes during peak activity times, demonstrating the robustness of the system in dynamic environments like mechanical workshops.

Compared to traditional monitoring systems, which rely on manual measurements or periodic analyses, the developed system proved to be much more efficient, providing continuous information and enabling immediate interventions when necessary. The ability to generate historical reports and identify long-term patterns also stands out as a significant improvement over less automated approaches. This is consistent with the literature, which highlights the effectiveness of IoT systems in industrial environments, especially in terms of process optimization and increased occupational safety.

Machine learning-based modeling, using algorithms such as Linear Regression, Decision Trees, and SVM, was crucial for analyzing the collected environmental variables. However, the main methodological contribution of the project was the use of Gaussian functions to model the relationship between temperature, humidity, and air quality. This approach proved effective in capturing the complexity of the interactions between these variables and providing an intuitive healthiness index, easily understood by workshop managers.

Although the use of Gaussian functions was effective, a limitation encountered was the need for continuous sensor calibration, especially in environments with variable ventilation. The literature points out that, in industrial environments, external factors such as changes in humidity and variations in air circulation can influence sensor accuracy. This may require periodic adjustments to the model to maintain prediction accuracy. Nevertheless, the results indicate that, when properly calibrated, the AI system provides an accurate and practical representation of air quality, aiding in real-time decision-making.

One of the most relevant points discussed in this project is the positive impact of the monitoring system on worker health and safety. The ability to issue automatic alerts in critical situations allowed workshops to take immediate preventive measures, reducing workers' exposure to dangerous levels of toxic gases and other pollutants. Implementing this type of technology has the potential to reduce the incidence of respiratory diseases and other health problems related to poor air quality, as discussed in studies on work environments in industrial sectors.

Although the system demonstrated robust performance, some limitations were observed. The need for constant sensor calibration, particularly in workshops with variable ventilation, was a challenge throughout the project. Another limitation was the dependency on internet connectivity for real-time data transmission. In areas with limited network infrastructure, data transmission latency may compromise the effectiveness of continuous monitoring.

Moreover, despite the accuracy of the AI algorithms used, the system could be improved by integrating more advanced methods, such as neural networks or deep learning algorithms, which could further refine air quality modeling and predict critical situations with greater anticipation. These algorithms could be trained based on more robust historical data, providing more accurate predictions of air pollution events.

Based on the results and the limitations observed, several directions for future developments are suggested. First, integrating additional sensors, such as PM2.5 and PM10 particle detectors, could provide a more comprehensive view of air quality in automotive workshops. Additionally, the use of more advanced AI algorithms, such as convolutional neural networks, could further enhance the system's analysis and prediction capabilities.

Another aspect that deserves attention is improving the system's user interface to make it even more intuitive and accessible, especially for users with little experience in digital technologies. Enhancing connectivity, such as adopting mesh networks or alternative connectivity technologies, could also be explored to ensure continuous data transmission, even in locations with limited network infrastructure.

\end{document}